\documentclass[10pt, journal]{IEEEtran}
\IEEEoverridecommandlockouts
\usepackage{cite}
\usepackage{amsmath,amssymb,amsfonts}
\usepackage{bm}
\usepackage{graphicx}
\usepackage{textcomp}
\usepackage{xcolor}
\usepackage[center]{caption}
\usepackage{amsthm}
\usepackage{bbm}
\usepackage{subfig}

\makeatletter
\newif\if@restonecol
\makeatother

\usepackage[linesnumbered,ruled,vlined]{algorithm2e}
\usepackage{algpseudocode}
\usepackage{amsmath}

\def\BibTeX{{\rm B\kern-.05em{\sc i\kern-.025em b}\kern-.08em
    T\kern-.1667em\lower.7ex\hbox{E}\kern-.125emX}}
\begin{document}

\title{Decentralized Federated Learning for UAV Networks: Architecture, Challenges, and Opportunities}
%

\author{Yuben~Qu,
        Haipeng~Dai,
        Yan~Zhuang,
        Jiafa~Chen,
        Chao~Dong,
        Fan~Wu,
        Song~Guo
\thanks{Y. Qu, J. Chen and C. Dong are with Nanjing University of Aeronautics and Astronautics, Nanjing, China. Y. Qu is also with Shanghai Jiao Tong University, Shanghai, China.}
\thanks{Y. Zhuang, and F. Wu are with Shanghai Jiao Tong University, Shanghai, China.}
\thanks{H. Dai is with Nanjing University, Nanjing, China.}

\thanks{S. Guo is with The Hong Kong Polytechnic University, Hong Kong, China.}
}

\maketitle

\begin{abstract}
Unmanned aerial vehicles (UAVs), or say drones, are envisioned to support extensive applications in next-generation wireless networks in both civil and military fields. Empowering UAVs networks intelligence by artificial intelligence (AI) especially machine learning (ML) techniques is inevitable and appealing to enable the aforementioned applications. To solve the problems of traditional cloud-centric ML for UAV networks such as privacy concern, unacceptable latency, and resource burden, a distributed ML technique, \textit{i.e.}, federated learning (FL), has been recently proposed to enable multiple UAVs to collaboratively train ML model without letting out raw data. However, almost all existing FL paradigms are still centralized, \textit{i.e.}, a central entity is in charge of ML model aggregation and fusion over the whole network, which could result in the issue of a single point of failure and are inappropriate to UAV networks with both unreliable nodes and links. Thus motivated, in this article, we propose a novel architecture called DFL-UN (\underline{D}ecentralized \underline{F}ederated \underline{L}earning for \underline{U}AV \underline{N}etworks), which enables FL within UAV networks without a central entity. We also conduct a preliminary simulation study to validate the feasibility and effectiveness of the DFL-UN architecture. Finally, we discuss the main challenges and potential research directions in the DFL-UN.
\end{abstract}

\section{Introduction}
Unmanned aerial vehicles (UAVs), also known as drones, are expected to play a critical role in numerous stirring applications in next-generation wireless networks, ranging from delivery of goods, monitoring, surveillance, to telecommunications in both civil and military fields. On one hand, owing to their flexibility, line-of-sight (LoS) connections, and 3D mobility, UAVs could act as flying base stations (BSs) to deliver communication/computing/caching services in future wireless networks, which compensates the shortcoming of traditional infrastructure-based networks. On the other hand, UAVs can also serve as flying users to support emerging applications including remote sensing, item delivery, target recognition/tracking, and even virtual reality. To enable the aforementioned applications, it is an inevitable and appealing trend to intelligentize UAV networks by artificial intelligence (AI) especially machine learning (ML) techniques.

While ML has already demonstrated its power to bring intelligence to wireless networks including UAV networks, traditional ML approaches are cloud-centric, \textit{i.e.}, all the data is required to be transmitted to a cloud data center and processed therein, which may be inappropriate for UAV networks \cite{FLU2020Brik}. Firstly, the data generated by each UAV may be inaccessible due to the well-known privacy concern, since it might contain some sensitive information (\textit{e.g.}, UAV's identity and localization). Secondly, the experienced latency from sending raw data to receiving well trained model is unacceptable for some real-time UAV applications (\textit{e.g.}, autonomous drones monitoring and target tracking). Lastly, the transfer of huge raw data such as image and video to the cloud consumes a lot of bandwidth and energy, which is unacceptable for UAV networks with limited bandwidth and energy supply. Hence, it would be greatly beneficial if the ML model training could be conducted in a distributive manner in UAV networks directly, without sending data out.

Recently, federated learning (FL), firstly proposed by Google \cite{CELD}, emerges as a promising distributed ML paradigm to solve the drawback of traditional cloud-centric ML. At its core, FL enables multiple devices to collaboratively train an ML model without sending the raw data out, thereby protecting device privacy, improving experienced latency, and relieving bandwidth and energy burden. It was demonstrated that FL is more suitable for wireless edge networks as compared to the cloud-centric ML \cite{Samarakoon2018FLUR}, because it enables wireless edge devices to collaboratively learn a shared ML model in parallel, while keeping all raw data on device. Furthermore, to better adapt to the characteristics of wireless edge networks such as multi-hop, several FL paradigms including collaborative FL \cite{Chen2020WCCFL}, multi-hop FL \cite{Pinya2020FedAir}, and fog learning \cite{Seyye2020FogL} have been proposed, while a first FL framework within UAV networks is recently presented in \cite{Zeng2020FLS}. Nevertheless, all the aforementioned FL schemes are \textit{centralized}, \textit{i.e.}, relying on a central entity for continuous ML model aggregation. They could face the problem of a single point of failure, which is thus inappropriate for UAV networks with unreliable nodes and links, \textit{i.e.}, when the UAV serving as the central entity is out of battery or the wireless link between it and other UAVs fails, the FL training would have to terminate. Worthy noting that there exist some studies about decentralized ML such as \cite{Ram2009AGA, Nedic2009DSM, Lian2017CDA, Tang2020CED}, where \cite{Ram2009AGA, Nedic2009DSM} fall into the scope of classical distributed ML rather than FL, and they seldom discuss how to apply decentralized FL to UAV networks.
\begin{figure*}[t]
\centering
\includegraphics[scale = 0.425]{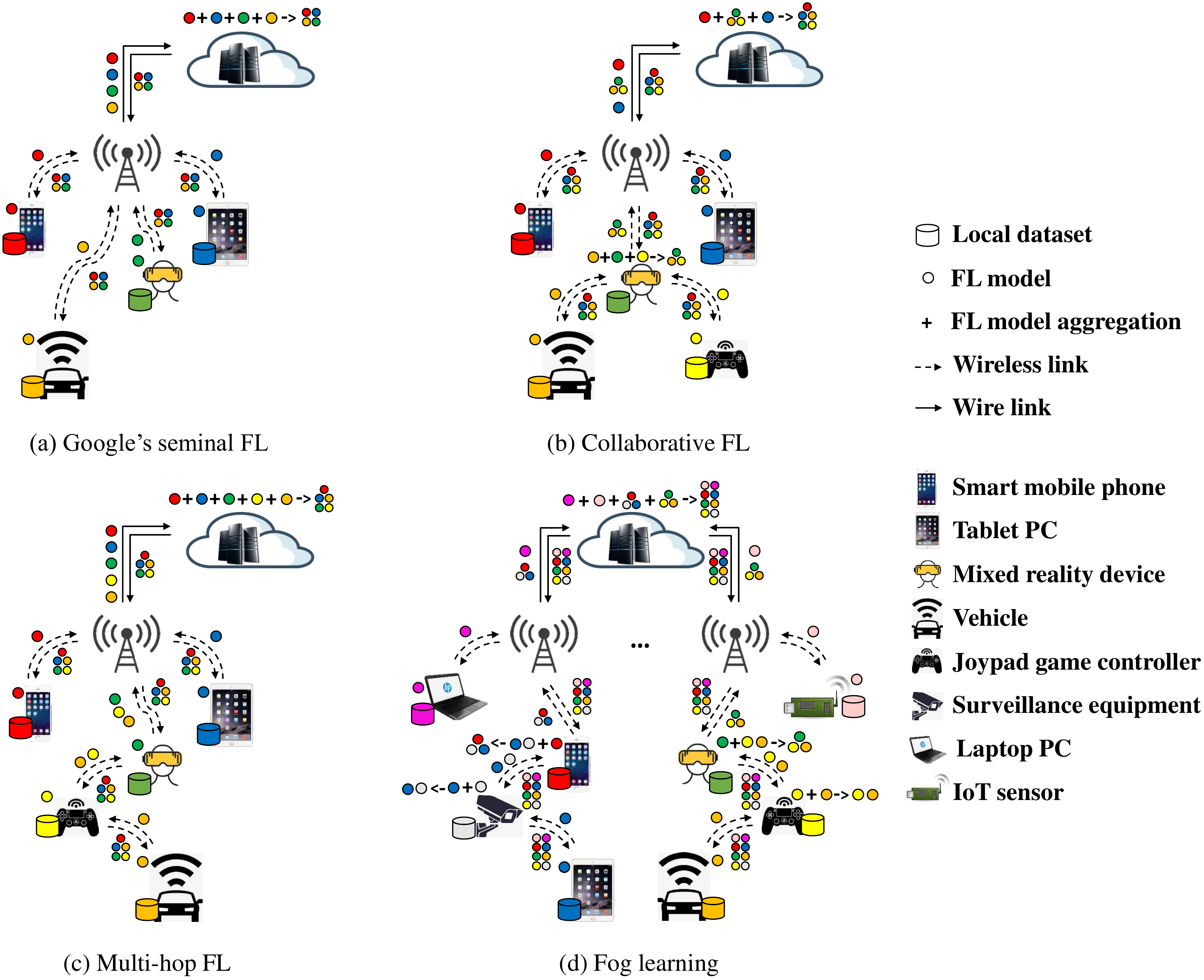}
\caption{Conventional FL and its representative counterparts.}
\label{CMFL}
\end{figure*}

In this article, to deal with the limitation of conventional FL schemes, we propose a novel architecture of \underline{D}ecentralized \underline{F}ederated \underline{L}earning for \underline{U}AV \underline{N}etworks (DFL-UN). Unlike conventional server-based FL schemes, while the DFL-UN follows the basic principles of FL (\textit{e.g.}, each UAV trains a local model based on its own data), it does not need a central entity for global model aggregation and fusion, that is to say, each UAV only exchanges local models with its one-hop neighboring UAVs by aggregation. To the best of our knowledge, this is first work that proposes a server-less FL architecture for UAV networks. We also present some preliminary results to validate the feasibility and effectiveness of the proposed DFL-UN, compared to a conventional FL scheme. In the rest of this article, we first introduce several conventional FL paradigms in detail and discuss the limitation for UAV networks in Section~II. The proposed DFL-UN architecture is formally presented in Section~III, whose effectiveness is validated by simulation results in Section~IV. In Section~V, we highlight several challenges and potential research directions in the DFL-UN. Section~VI concludes this article.

\section{Conventional FL Paradigms and Its Limitation for UAV Networks}
In this section, we first introduce several conventional FL paradigms including the original FL and its three representative counterparts, and then discuss their main limitation for UAV networks.
\begin{figure*}[t]
\centering
\includegraphics[scale = 0.4]{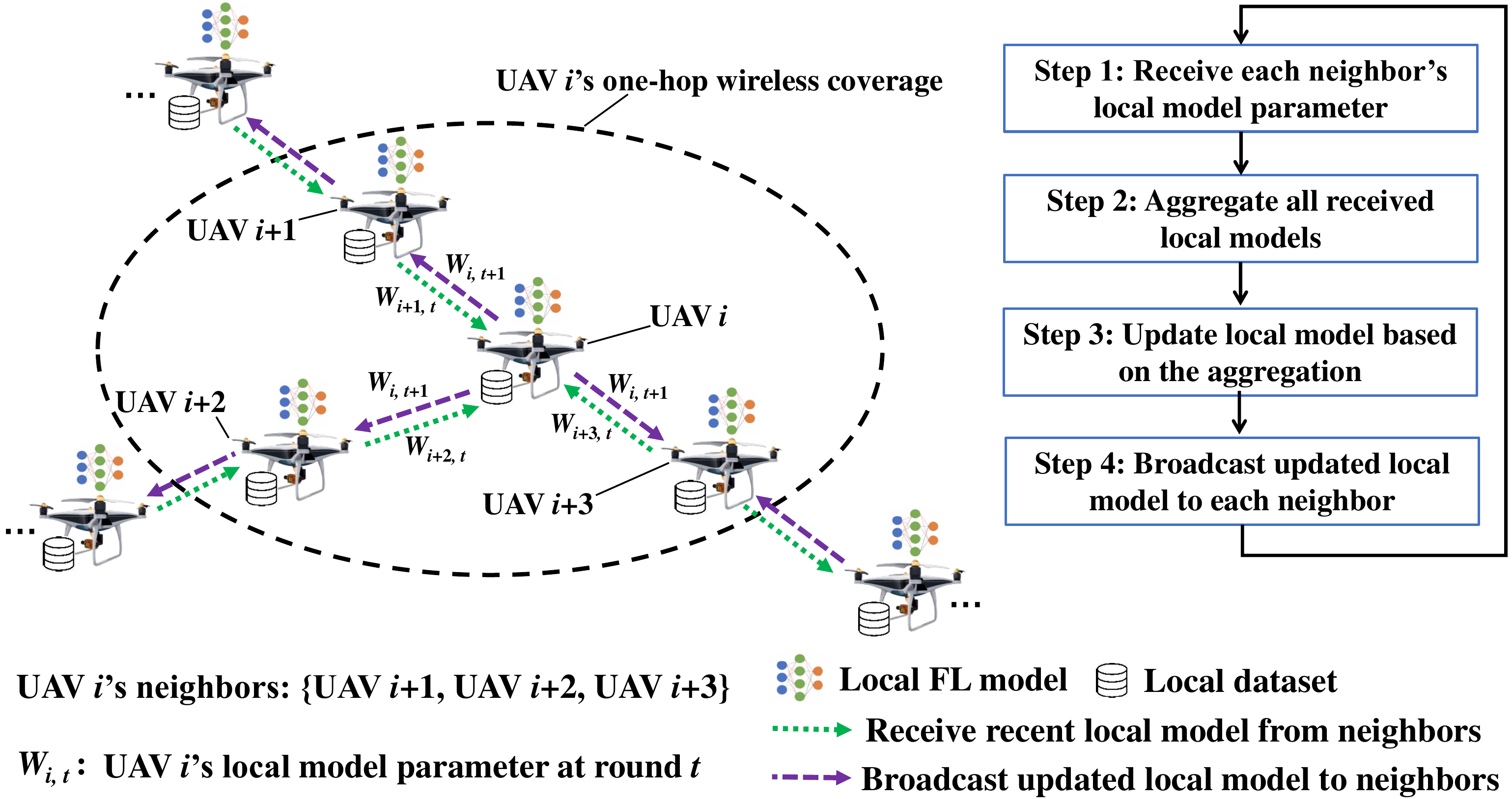}
\caption{An illustration of the proposed DFL-UN architecture for UAV networks.}
\label{SELF-UN_Illustration}
\end{figure*}
\subsection{Overview of Conventional FL Paradigms}

\textbf{Google's seminal FL:} to reduce privacy risk and communication cost, McMahan \textit{et al.} \cite{CELD} proposed the first FL framework. Different from the centralized ML, FL enables each user equipment (UE) to train a local ML model based on its local dataset rather than directly send out its own data, which is transmitted to a cloud (called ``parameter server'') for global model averaging, as shown in Fig.~\ref{CMFL}~(a). Note that the FL model is trained in an iterative manner, \textit{i.e.}, the local model is always updated based on the latest averaged global model, until the convergence about the global model is achieved. Recently, much efforts have been devoted to \textit{wireless edge FL}, where the global model averaging happens at the one-hop wireless edge server (\textit{e.g.}, \cite{Chen2021JLCF,Tran2019FLWN,Ren2021ADNN}), rather than the remote cloud server.

\textbf{Collaborative FL:} Chen \textit{et al.} \cite{Chen2020WCCFL} proposed the concept of collaborative FL, where some UEs far away from, or even not connected to a cloud or a BS can engage in FL by device-to-device (D2D) communications, \textit{i.e.}, transmitting their local models to nearby neighbors associated with the BS, as illustrated in Fig.~\ref{CMFL}~(b). Note that in collaborative FL, any UE should aggregate its received local models from all neighbors and then send the aggregated local model to the BS. While Google's seminal FL can been seen as a special case of collaborative FL, \textit{i.e.}, they are equivalent when all UEs are connected to a BS, the prominent advantage of the latter is that it can involve more UEs and data for better training performance \cite{Chen2020WCCFL}.

\textbf{Multi-hop FL:} similar to collaborative FL involving more UEs for training, Pinyoanuntapong \textit{et al.} \cite{Pinya2020FedAir} proposed multi-hop FL, which aims to enable FL over wireless multi-hop networks. In multi-hop FL, the local models of some UEs not directly connected to the parameter server node will be forwarded by the UE based on the routing policy, as shown in Fig.~\ref{CMFL}~(c). The only difference between collaborative FL and multi-hop FL is that the local models received from neighbors are aggregated before forwarding by each UE in the former, while they are directly forwarded in the latter.

\textbf{Fog learning:} considering both network and topology structures in fog environment, Hosseinalipour \textit{et al.} \cite{Seyye2020FogL} proposed fog learning, which intelligently distributes model training across the continuum of various nodes from edge UEs to cloud servers, as illustrated in Fig.~\ref{CMFL}~(d). Similar to collaborative ML, fog learning employs D2D links to orchestrate heterogeneous UEs with various proximities, which forms a multi-layer hybrid FL framework. While UEs also aggregate the local models received from neighbors in fog learning, it divides all UEs into different layers according to their proximities, which are not clearly divided in collaborative FL.

\subsection{Limitation for UAV Networks}
Although the aforementioned collaborative FL, multi-hop FL, and fog learning have modified the celebrated Google's seminal FL to migrate from the star network topology for model parameters exchange in FL to more distributed topology at scale, all of them are \textit{centralized}, \textit{i.e.}, they rely on a central parameter server for global model aggregation and fusion. Unlike the relatively reliable parameter server (\textit{e.g.}, cellular BS) in terrestrial FL, such server-based FL is inevitably faced with a single point of failure when applied in UAV networks. For example, in the FL framework within a swarm of multiple wireless-connected UAVs proposed in \cite{Zeng2020FLS}, a leading UAV acts as the parameter server, while several following UAVs train their local models and send them to the leading UAV for aggregation. The leading UAV may be unreachable since air-to-air (A2A) wireless links are subject to errors, or even cannot work properly due to attacks or out of battery.
\begin{figure*}[t]
\centering
\includegraphics[scale = 0.4]{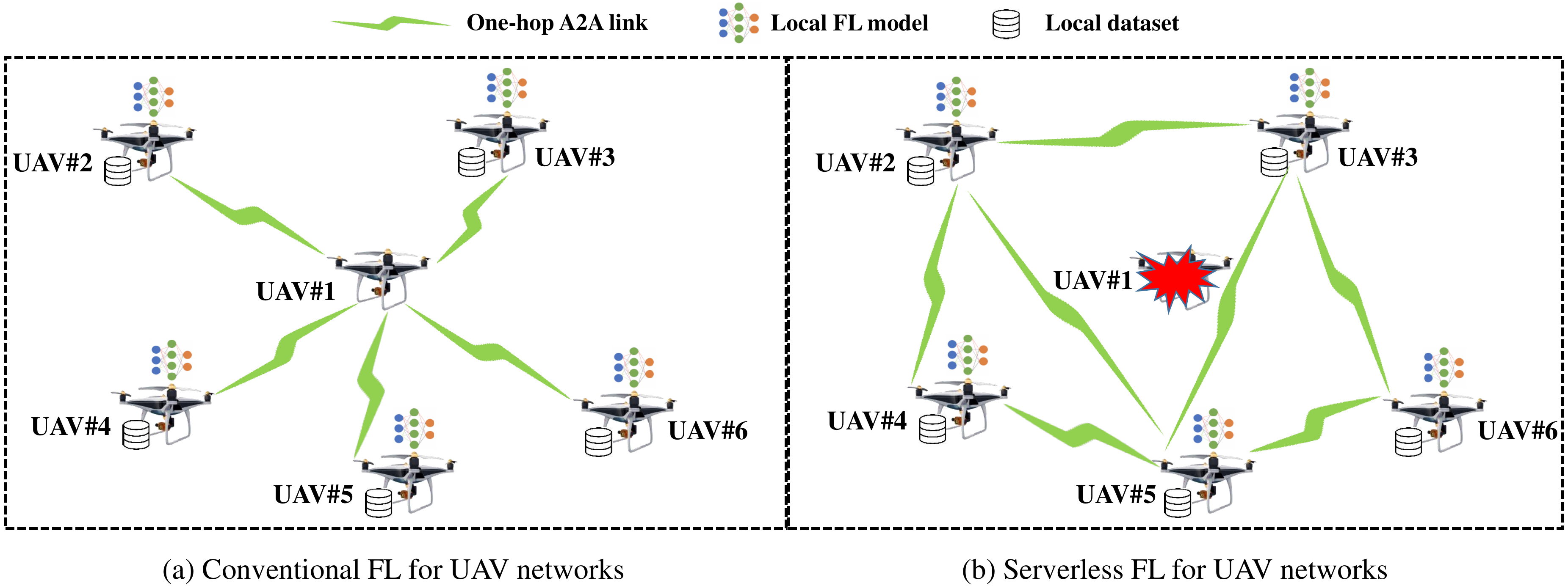}
\caption{The UAV network topology in the simulation.}
\label{SimulationNetwork}
\end{figure*}
\section{Decentralized FL for UAV Networks: Architecture, Advantages, and Novelty}
In this section, to resolve the aforementioned limitation of conventional FL, we propose the architecture of \underline{D}ecentralized \underline{F}ederated \underline{L}earning for \underline{U}AV \underline{N}etworks (DFL-UN). In the following, we first present the architecture overview of the DFL-UN, and then explain its advantages and novelty.

\textbf{Architecture:} in general, in the proposed DFL-UN architecture, the FL is conducted over different UAVs in a fully distributed manner, without a central parameter server for global model aggregation across the UAV network. More specifically, each UAV always employs its local dataset to update a local model, upon receiving the local model weights from its neighboring UAVs. Take Fig.~\ref{SELF-UN_Illustration} as an example. Assume that UAV $i$'s one-hop neighbors are UAVs $i+1$, $i+2$, and $i+3$. Each UAV trains a local FL model based on its local dataset, where $W_{i,t}$ represents UAV $i$'s local model parameter (weight) at training round $t$. We suppose that the data are respectively collected by the UAVs and labeled based on the local observation according to the given phenomenon or available public UAV datasets. Initially, each UAV $i$ generates a local model $W_{i, 0}$. Firstly, at round $t$, UAVs $i+1$, $i+2$, and $i+3$ send their corresponding model weights $W_{i+1, t}$, $W_{i+2, t}$, and $W_{i+3, t}$ to UAV $i$. Secondly, these neighbors' model weights and UAV $i$'s current model weight are used to generate an aggregated local model at UAV $i$. Thirdly, based on that aggregated local model, UAV $i$ updates its local model as $W_{i, t+1}$, which will be ``broadcast'' to the neighbors afterwards for aggregation and update at each neighbor. In essence, while each UAV acts as a ``parameter server'' to aggregate local models for its neighbors, it also updates its own local model based on the aggregated local model.

Note that to avoid large latency and low efficiency by full wireless broadcasting, we suppose that the model parameter exchange between any two neighboring UAVs is transmitted by D2D communications, and the channel access among these D2D pairs could be efficiently managed by using a Time Division Multiple Access (TDMA) or a Frequency Division Multiple Access (FDMA) medium sharing scheme. How to optimize D2D communications with resource allocation to boost the learning performance of the DFL-UN is left for future study. By this way, the information of each local model is in fact propagated over the UAV network by multiple D2D exchanges. Similar to \cite{Tang2020CED}, there exists a coordinator in the proposed DFL-UN, who is in charge of initializing the FL training task, distributing the global information to all involved UAVs, and monitoring the training process of the FL.

\textbf{Advantages and novelty:} the proposed DFL-UN architecture can well adapt to UAV networks with high dynamics in both A2A wireless links and network topology. This brings two main advantages when boosting edge/on-device intelligence for UAV networks as follows. i) The DFL-UN can enable distributed ML with high robustness over UAV networks. Specifically, since there exists no such a central node coordinating the learning process in the DFL-UN, the FL will not terminate if any UAV or A2A link is unavailable. ii) The DFL-UN provides high flexibility and agility for collaborative ML within UAV networks. No matter how the network topology changes due to the dynamic joining and leaving of some UAVs, the FL does not need to reorganize and will continue with negligible efforts. In a nutshell, the novelty of the DFL-UN lies in that it proposes a fully decentralized FL framework for UAV networks, which perfectly matches the unique characteristics of UAV networks.

\section{Performance Evaluation}
In this section, we conduct numerical simulations to validate the feasibility as well as effectiveness of the proposed DFL-UN architecture. We study the performance of the DFL-UN in a UAV network, where one of the UAVs acting as the parameter server is unavailable during the training period. We involve the conventional centralized FL (\textit{i.e.}, FedAvg \cite{CELD}) as the benchmark, where the parameter server node is assumed to be available in this case. Therefore, the learning performance of the conventional FL could be seen as a theoretical upper bound for the DFL-UN. The DFL-UN is effective if the gap of the performances between it and conventional FL is very small after multiple training rounds. Here a training (communication) round is defined as follows: the training will not enter the next communication round until all UAVs have completed the four steps in Fig.~2.

\subsection{Simulation Settings}
We assume that there exist six UAVs flying at the fixed altitude of 100 m in a UAV network, five of which aim to collaboratively train an ML model by FL based on their own data. The distance between each two UAVs is randomly chosen from [80, 120] m. As shown in Fig.~\ref{SimulationNetwork}, if UAV\#1 who acts as the parameter server is available (UAV\#1 can directly communicate with the rest UAVs), the rest five UAVs can conduct the conventional centralized FL with the help of UAV\#1 (Fig.~\ref{SimulationNetwork}~(a)); otherwise, they can only conduct the decentralized FL following the proposed DFL-UN (Fig.~\ref{SimulationNetwork}~(b), the link relationship among the UAVs is also illustrated in this figure). The learning rate is fixed at 0.025 and the number of local epochs is set as 3 with the mini-batch size 5. In the simulation, for the training task, we consider the classification task using Convolutional Neural Network (CNN) model \cite{Savazzi2016DFHS}. For the training dataset, we assume each UAV collects 25 samples and the data distribution is non-IID, and the number of CPU cycles needed per sample is 6$\times10^4$. The data size of the FL model parameter is 56 Kbits. For the setting of the UAV network, in the communication, we assume the channel power gain, noise power, the transmission power, and spectrum bandwidth allocated to each UAV, as -50 dBm, -90 dBm, 30 dBm, and 0.4 MHz, respectively; in the computation, the computing capacity of each UAV is randomly selected from [1, 2] GHz. For the performance metrics, we consider the cross-entropy loss (including both the average loss of all UAVs during the training period and individual loss of each UAV after the training), and the overall training latency, which mainly consists of three parts: i) the latency for receiving the local model parameter from each neighbor; ii) the latency for aggregating all received local models and updating local model based on the aggregation; iii) the latency for transmitting the updated local model to each neighbor.
\subsection{Result Analysis}
\begin{figure*}[t]
\centering
\subfloat[Average loss]{\includegraphics[height=1.5in]{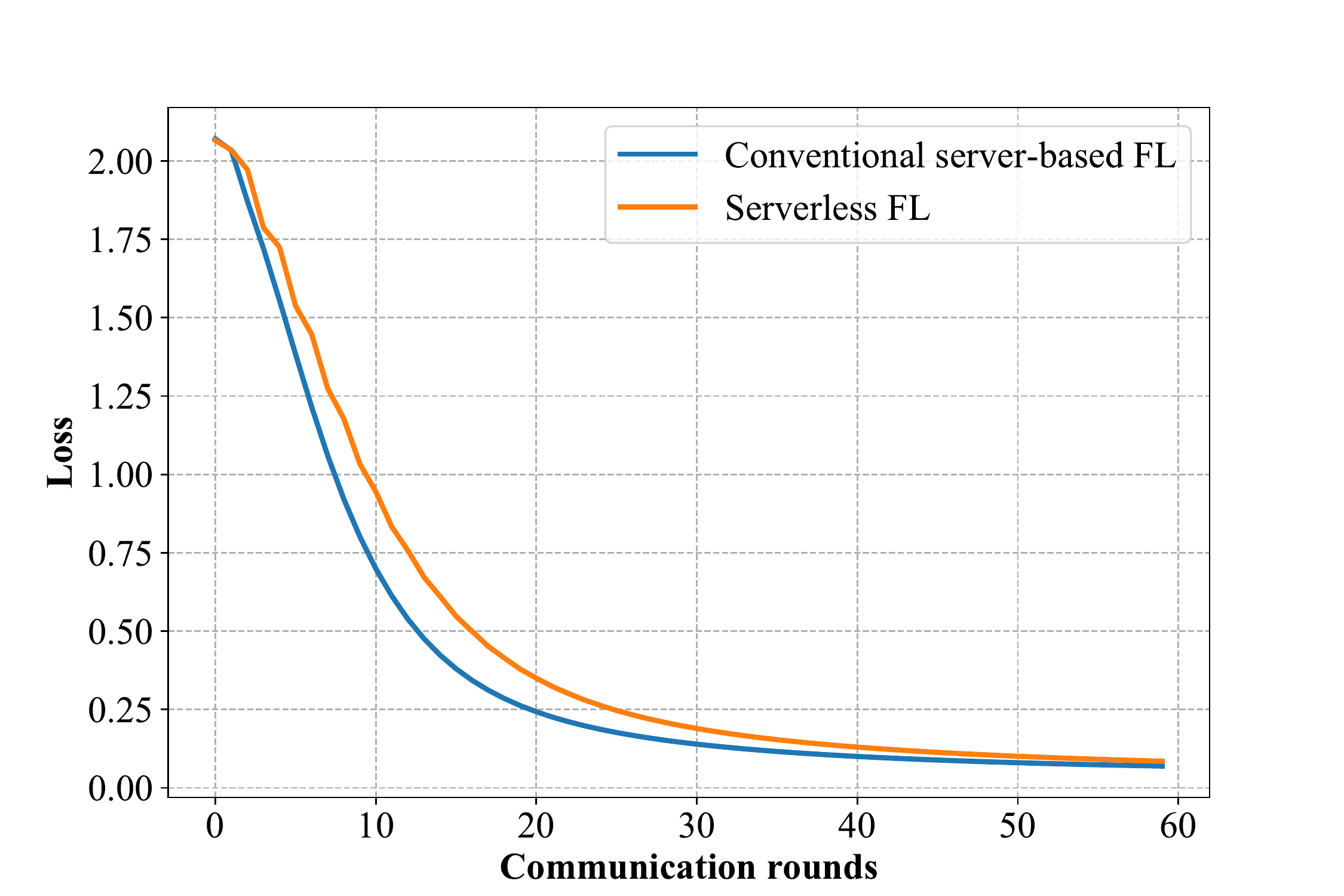}}\hspace{0.1cm}
\subfloat[Individual loss]{\includegraphics[height=1.5in]{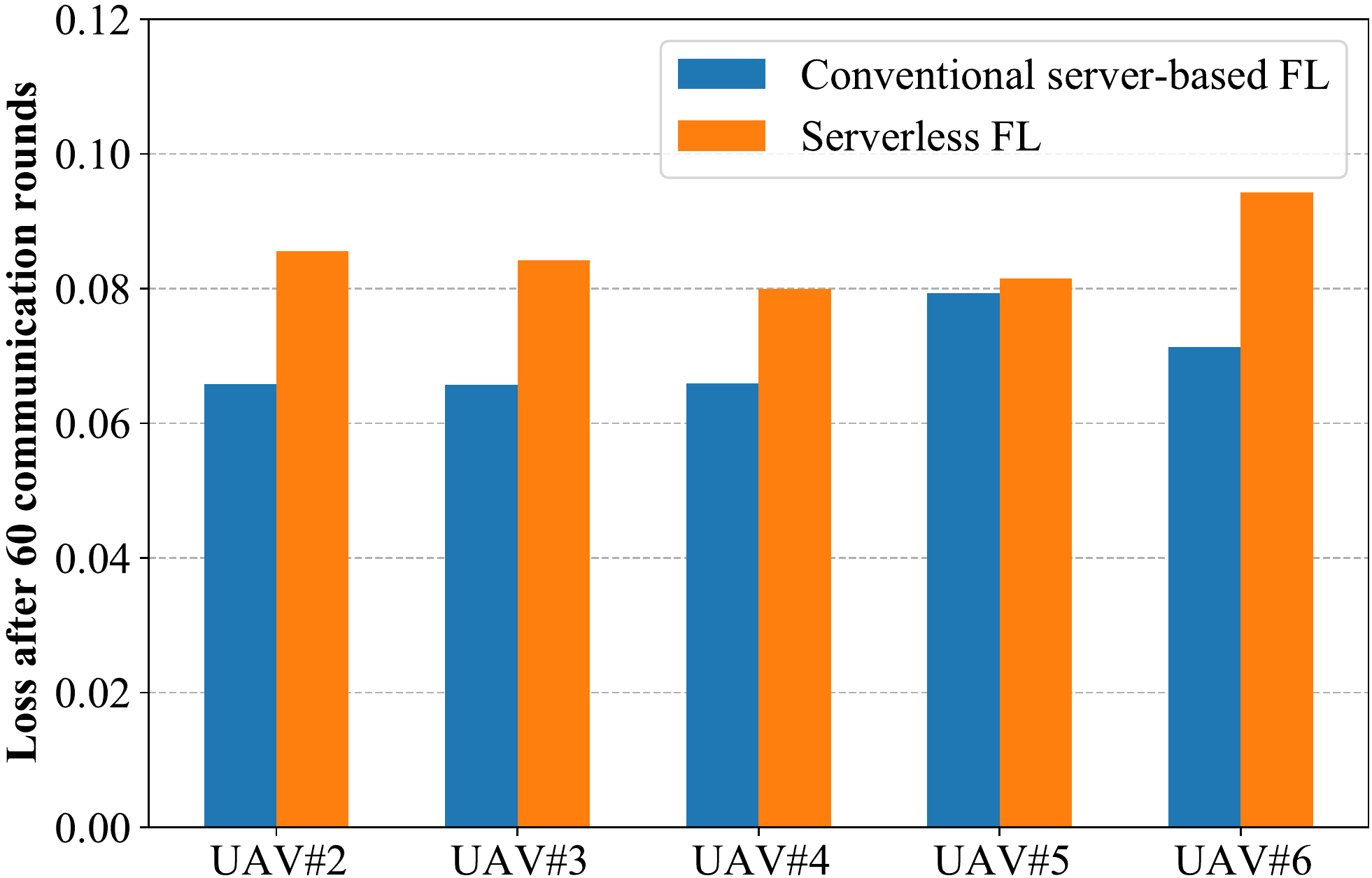}}\hspace{0.1cm}
\subfloat[Training latency]{\includegraphics[height=1.5in]{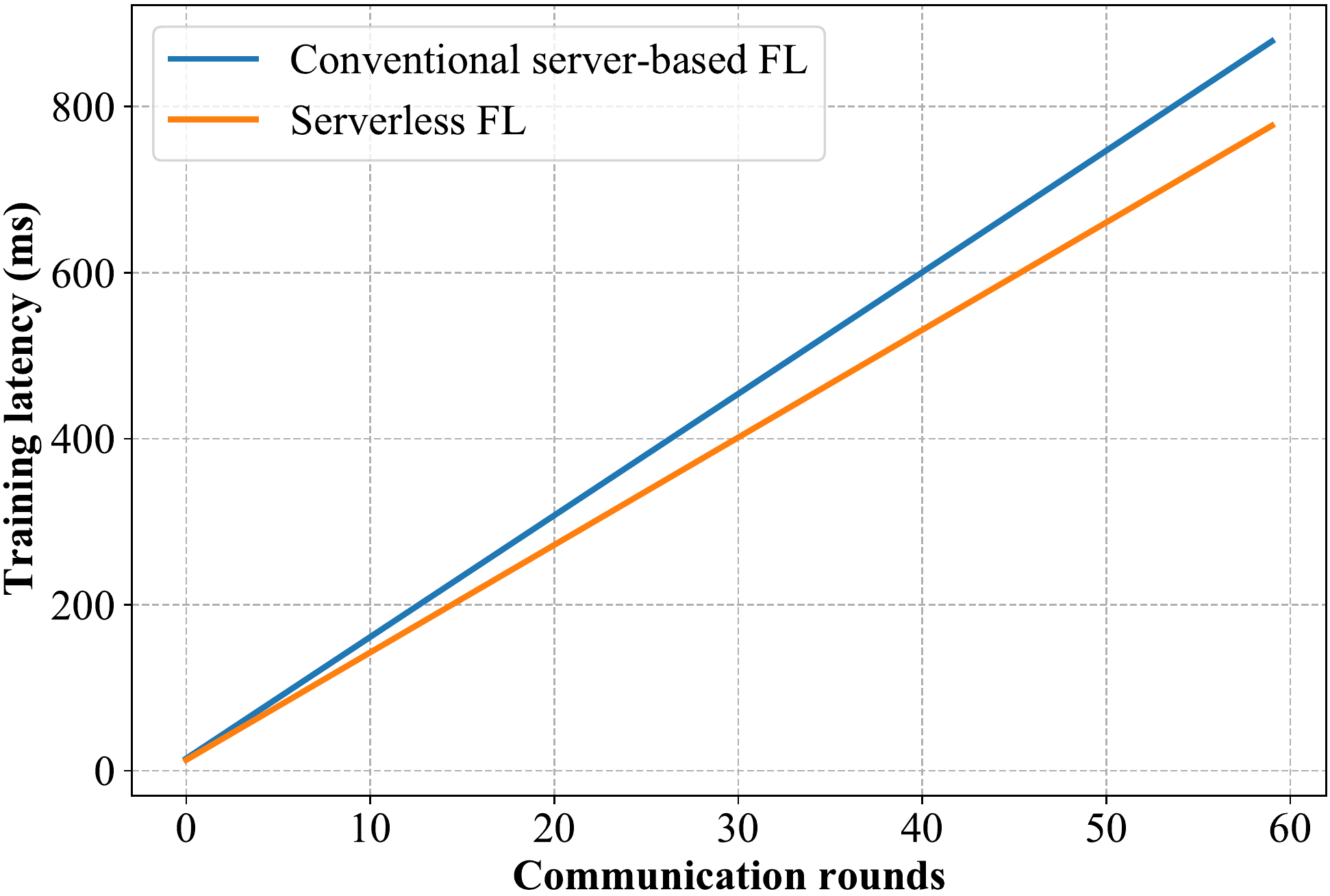}}
\caption{Evaluation results of the decentralized FL compared with conventional centralized FL.}
\label{Results}
\end{figure*}

Fig.~\ref{Results} presents the performance evaluation results of the decentralized FL in the proposed DFL-UN and conventional centralized FL for the UAV network in Fig.~\ref{SimulationNetwork}. Specifically, as shown in Fig.~\ref{Results}~(a), although the average loss value achieved by the decentralized FL is always higher than that by the conventional centralized FL during the training, the final gap between them is only 0.0156, which implies that the decentralized FL is almost as effective as the conventional centralized FL, without the help of a central parameter server. Furthermore, according to Fig.~\ref{Results}~(b), for each UAV, the difference of the loss value after 60 communication rounds between the decentralized FL and the conventional centralized FL is no more than 0.0229 only. And the individual loss of each UAV after 60 rounds is totally different owing to the following reasons. Firstly, since there is no such a global FL model trained in the proposed DFL-UN, the final local FL model will not be updated based on a shared global FL model and differs at different UAV. Secondly, the data at each UAV is also different. Thus, based on the definition of the (cross-entropy) loss, the individual loss of each UAV is different. Lastly, as illustrated in Fig.~\ref{Results}~(c), the training latency by the decentralized FL is always smaller than that by the conventional centralized FL, since in the latter it needs to broadcast the global aggregated FL model in each communication round, where the final decreased latency is about 101.72 ms. We also notice that the training latency curves of both schemes are linear to the number of communication rounds, mainly because both the network topology and resource allocation within each communication round are identical, thereby making the training latency within each round identical. To sum up, our results validate the effectiveness of the proposed DFL-UN, in terms of achieving almost the same learning performance with lower training latency. Note that since it is only a preliminary simulation study, we would like to conduct more extensive evaluation studies in the future, \textit{e.g.}, the inference accuracy of the trained model by the DFL-UN, and robustness to the failures of the UAVs in the DFL-UN.

\section{Challenges and Potential Research Directions}
\subsection{Main Technical Challenges}
The proposed DFL-UN can avoid the drawback of single point of failure in the distributed collaborative ML, especially when some UAV acting as the parameter server fails or is unreachable. And the effectiveness of the DFL-UN compared to conventional FL is also validated by the preliminary evaluation results in Section~IV. However, there exist many technical challenges in the DFL-UN to be addressed before fully realizing its benefits as follows.

\textbf{Convergence judgment vs. Fully distributed learning:} in conventional FLs, there is a central controller that monitors whether the whole training converges or not in a straightforward way. In the DFL-UN, the whole training is actually asynchronous and fully distributed over multiple UAVs. Although there also exists a coordinator to monitor the training process of the FL following \cite{Tang2020CED}, the communication link between it and each UAV might be temporary in the context of highly dynamic UAV networks, which makes it challenging to timely judge the convergence sometimes. 

\textbf{Learning robustness vs. UAV network unreliability and heterogeneity:} UAV networks are intrinsically unreliable since UAVs with high mobility are easily affected by environment factors such as mechanical and wind vibrations, and the A2A links are wireless and unreliable. And different UAVs are probably heterogeneous in terms of collected data samples for training, computation and communication capabilities, and available energy. How to guarantee the robustness of the learning in the DFL-UN facing the network unreliability and heterogeneity is extremely challenging.

\textbf{Communication and energy efficiency vs. Extensive A2A transmissions:} unlike conventional FLs that involve no D2D (\textit{i.e.}, A2A) data transmissions or a small amount of A2A transmissions, the DFL-UN incurs extensive A2A transmissions, which makes the issue of communication and energy efficiency more severe. Specifically, considering the limited energy of each UAV, if it spends much energy in the A2A communication for the local model exchange, the energy left for the local model computation may be in a shortage. The joint FL learning and spectrum sharing optimization problem is also worthy studied.

\subsection{Potential Research Directions}
In light of the aforementioned technical challenges as well as the development trend of future UAV networks, we discuss the following potential research directions in the future study.

\textbf{Convergence behavior analysis:} as mentioned in Section V-A, it is challenging to judge the convergence of the DFL-UN. Intuitively, the convergence behavior of the DFL-UN not only relates to the evolution of this fully distributed collaborative ML, which is totally different from conventional FLs, but also is greatly affected by the unique characteristics of UAV networks such as dynamic topology and intermittent A2A links. In fact, the analysis of the convergence behavior of the DFL-UN is an open problem so far.

\textbf{Secure model exchange by blockchain:} similar to conventional FLs, the DFL-UN involves extensive model parameters data exchange among the UAVs therein, which may be exploited by malicious participants. For the data security of the DFL-UN, blockchain that is also distributed could be leveraged to enable the secure model exchange in FL, in the presence of malicious UAVs. Nonetheless, the blockchain-based DFL-UN might incur extra significant overhead including additional block propagation, which deserves deeply investigation.

\textbf{Energy/communication-efficient FL strategy:} the DFL-UN is different from conventional FLs as well as recent FLs over general wireless networks \cite{Chen2021JLCF, Tran2019FLWN}, which makes existing energy/communication-efficient FL strategies not applicable. Specifically, from a role standpoint, compared to conventional FLs, each node (\textit{i.e.}, UAV) takes more responsibility, \textit{i.e.}, model aggregation besides local model training, which consumes more energy. And the sharing of communication resources within the DFL-UN is also essentially different.

\textbf{Joint UAV placement, network resource, and training parameters optimization:} motivated by the above challenges, a critical but challenging question is how to jointly optimize the placement of UAVs, various network resources, and basic training parameters, to achieve the considerable learning performance. Many critical factors including UAV network unreliability and heterogeneity should be considered in the optimization problem. Thus, unlike existing works such as \cite{Zeng2020FLS} optimizing network resources only or \cite{Chen2021JLCF} jointly optimizing network resources and learning parameters, the problem is brand-new and cannot be solved by existing UAV placement strategies as well as joint network resource and training optimization methods.

\textbf{Adaptive decentralized FL for clustering/hierarchical UAV networks:} in practical, many UAV networks are cluster-based or hierarchical rather than ad hoc, where several UAVs form a small group with a leader who is in charge of the rest members. In such UAV networks, the proposed DFL-UN could be modified to enhance the performance, \textit{e.g.}, borrowing the ideas in centralized FL such as \cite{Chen2020WCCFL, Seyye2020FogL}, each group could be seen as a virtual ``node'' that exchanges its local model parameter with other neighboring ``node'', while the local models within any group are always aggregated as a single local model of this group.
\section{Conclusion}
In this article, we have proposed the DFL-UN architecture to achieve fully decentralized FL over UAV networks. To be specific, each UAV not only trains a local FL model over its own data, but also aggregates several FL models from its neighboring UAVs for the update of the local FL model, which does not need a central entity for global model aggregation and fusion over the whole network. The key advantages lie in the high robustness of FL and flexibility and agility. We have also presented some preliminary results to validate its feasibility and effectiveness, and discussed several challenges and promising future research directions in the DFL-UN.

\bibliographystyle{IEEEtran}

\begin{IEEEbiography}[{\includegraphics[width=0.8in,height=1in,clip,keepaspectratio]{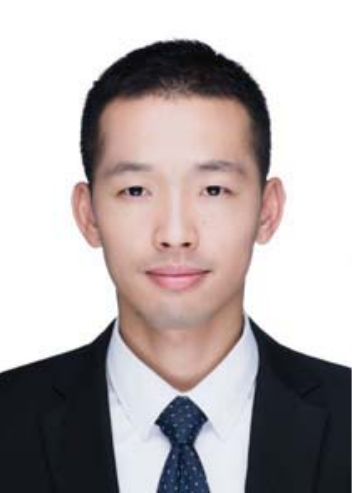}}]{Yuben Qu} received the B.S. degree in Mathematics and Applied Mathematics from Nanjing University, and both the M.S. degree in Communication and Information Systems and the Ph.D degree in Computer Science and Technology from Nanjing Institute of Communications, in 2009, 2012 and 2016, respectively. He is currently a research assistant in the College of Electronic and Information Engineering, Nanjing University of Aeronautics and Astronautics, Nanjing, China, and also a post-doc in the Department of Computer Science and Engineering, Shanghai Jiao Tong University, China. He is a recipient of The 2019 Post-doc Innovative Talent Support Program. From October 2015 to January 2016, he was a visiting research associate in the School of Computer Science and Engineering, The University of Aizu, Japan. His current research interests include edge intelligence computing, air-ground integrated networks, D2D communications, and crowdsensing.
\end{IEEEbiography}

\begin{IEEEbiography}[{\includegraphics [width=0.8in,height=1in,clip,keepaspectratio] {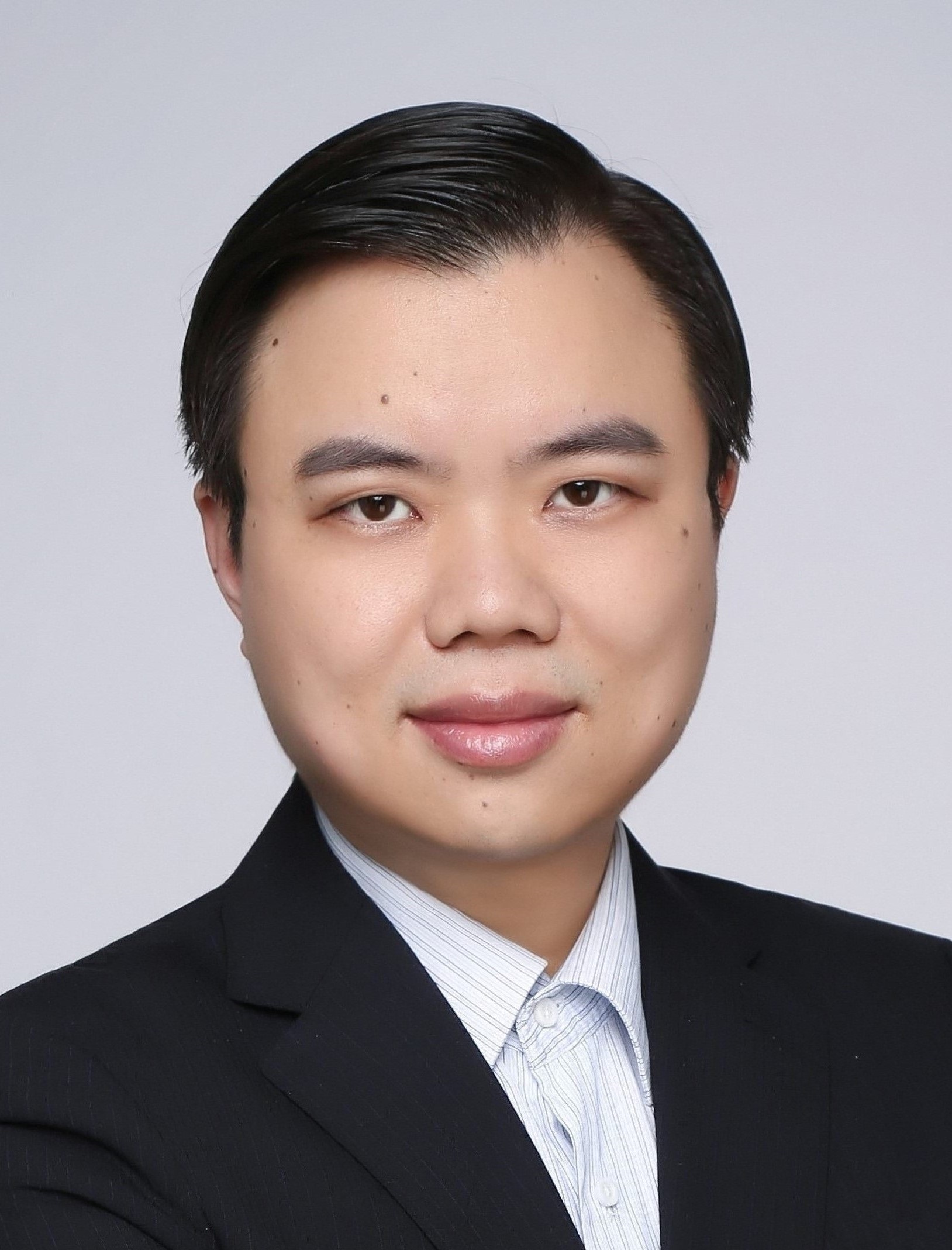}}]{Haipeng Dai}
received the B.S. degree in the Department of Electronic Engineering from Shanghai Jiao Tong University, Shanghai, China, in 2010, and the Ph.D. degree in the Department of Computer Science and Technology in Nanjing University, Nanjing, China, in 2014.
His research interests are mainly in the areas of wireless charging, mobile computing, and data mining.
He is an associate professor in the Department of Computer Science and Technology in Nanjing University.
His research papers have been published in many prestigious conferences and journals such as ACM MobiSys, ACM MobiHoc, ACM VLDB, IEEE ICDE, ACM SIGMETRICS, ACM UbiComp, IEEE INFOCOM, IEEE ICDCS, IEEE ICNP, IEEE SECON, IEEE IPSN, IEEE JSAC, IEEE/ACM TON, IEEE TMC, IEEE TPDS, and IEEE TOSN.
He is an IEEE and ACM member.
He serves/ed as Poster Chair of the IEEE ICNP'14, Track Chair of the ICCCN'19, TPC member of the ACM MobiHoc'20-21, IEEE INFOCOM'20-21, IEEE ICDCS'20-21, IEEE ICNP'14, IEEE IWQoS'19-21, IEEE IPDPS'20 and IEEE MASS'18-19.
He received Best Paper Award from IEEE ICNP'15, Best Paper Award Runner-up from IEEE SECON'18, and Best Paper Award Candidate from IEEE INFOCOM'17.
\end{IEEEbiography}

\begin{IEEEbiography}[{\includegraphics [width=0.8in,height=1in,clip,keepaspectratio] {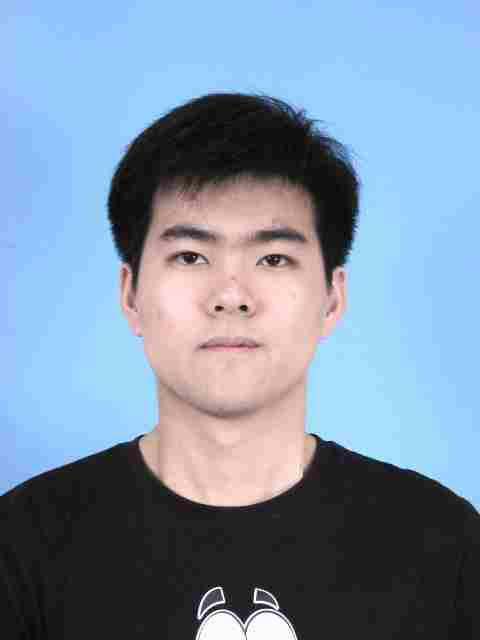}}]{Yan Zhuang} received his B.S. degree in the Department of Automation, University of Science and Technology of China, in 2019. He is currently working towards his M.S. degree in the Department of Computer Science and Engineering, Shanghai Jiao Tong University, China. His research interests include federated learning, wireless network and mobile edge computing.
\end{IEEEbiography}

\begin{IEEEbiography}[{\includegraphics [width=0.8in,height=1in,clip,keepaspectratio] {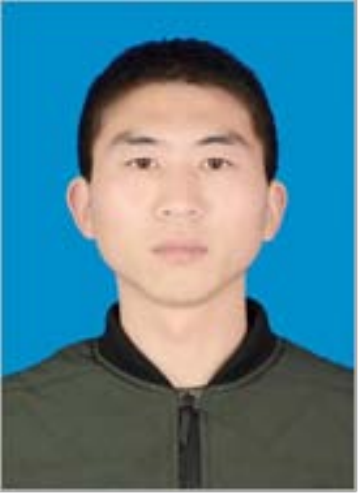}}]{Jiafa Chen} received the B.Eng. degree in communication engineering from Putian University, China, in 2017. He is currently a PhD student with the College of Electronic and Information Engineering, Nanjing University of Aeronautics and Astronautics, Nanjing, China. His research interests include energy harvesting communications, mobile edge computing, and radio resource management.
\end{IEEEbiography}

\begin{IEEEbiography}[{\includegraphics [width=0.8in,height=1in,clip,keepaspectratio] {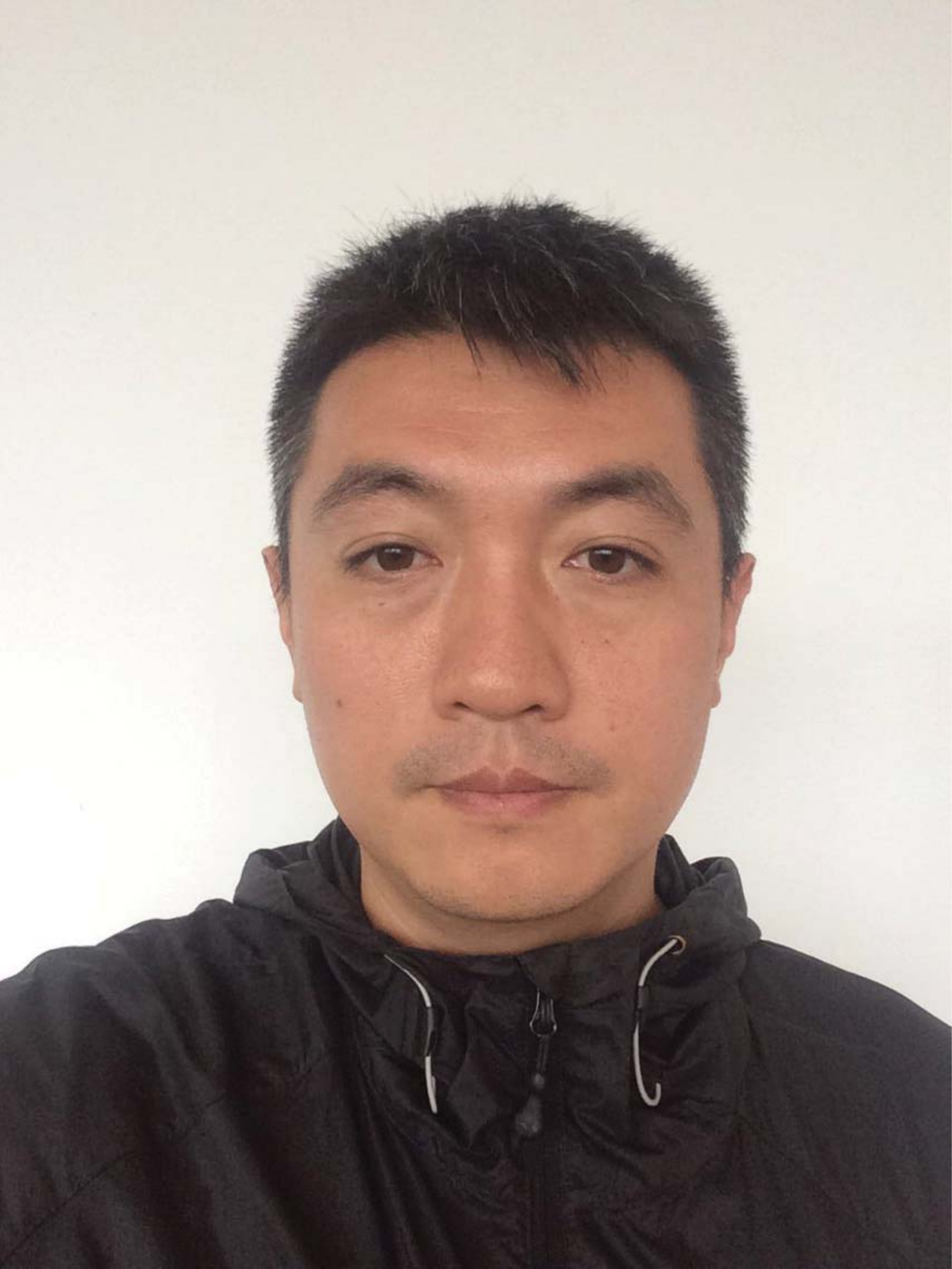}}]{Chao Dong} received his Ph.D degree in Communication Engineering from PLA University of Science and Technology, China, in 2007. From 2008 to 2011, he worked as a post Doc at the Department of Computer Science and Technology, Nanjing University, China. From 2011 to 2017, he was an Associate Professor with the Institute of Communications Engineering, PLA University of Science and Technology, Nanjing, China. He is now a full professor with the College of Electronic and Information Engineering, Nanjing University of Aeronautics and Astronautics, Nanjing, China. His current research interests include D2D communications, UAVs swarm networking and anti-jamming network protocol. He is a member of IEEE, ACM and IEICE.
\end{IEEEbiography}
\begin{IEEEbiography}[{\includegraphics [width=0.8in,height=1in,clip,keepaspectratio] {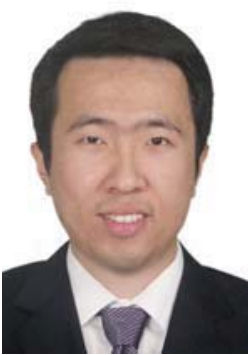}}]{Fan Wu} is a professor in the Department of Computer Science and Engineering, Shanghai Jiao Tong University. He received his B.S. in Computer Science from Nanjing University in 2004, and Ph.D. in Computer Science and Engineering from the State University of New York at Buffalo in 2009. His research interests include wireless networking and mobile computing, algorithmic game theory and its applications, and privacy preservation. He has published more than 150 peer-reviewed papers in technical journals and conference proceedings. He has served as an editor of IEEE Transactions on Mobile Computing, an area editor of Elsevier Computer Networks, and as the member of technical program committees of more than 90 academic conferences. For more information, please visit http://www.cs.sjtu.edu.cn/$\sim$fwu/.
\end{IEEEbiography}

\begin{IEEEbiography}[{\includegraphics[width=0.8in,height=1in,clip,keepaspectratio]{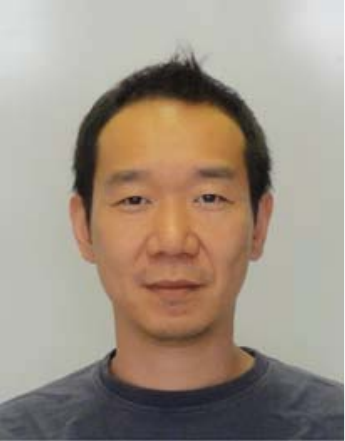}}]{Song Guo} is a Full Professor at Department of Computing, The Hong Kong Polytechnic University. His research interests are mainly in the areas of big data, cloud computing, mobile computing, and distributed systems with over 450 papers published in major conferences and journals. His work was recognized by the 2016 Annual Best of Computing: Notable Books and Articles in Computing in ACM Computing Reviews. He is the recipient of the 2018 IEEE TCGCC Best Magazine Paper Award, 2017 IEEE Systems Journal Annual Best Paper Award, and other six Best Paper Awards from IEEE/ACM conferences. Prof. Guo was an Associate Editor of IEEE Transactions on Parallel and Distributed Systems and an IEEE ComSoc Distinguished Lecturer. He is now an Associate Editor of IEEE Transactions on Cloud Computing, IEEE Transactions on Emerging Topics in Computing, IEEE Transactions on Sustainable Computing, IEEE Transactions on Green Communications and Networking, and IEEE Network. Prof. Guo also served as General and Program Chair for numerous IEEE conferences. He currently serves as a Director and Member of the Board of Governors of IEEE ComSoc. He is a fellow of the IEEE.
\end{IEEEbiography}

\end{document}